\documentclass[twocolumn,prd,nofootinbib,aps,floats,floatfix,amsmath,amssymb,secnumarabic]{revtex4} %
\usepackage[final]{graphicx}
\usepackage{amsmath}
\usepackage{bbm}
\usepackage{amsfonts}
\usepackage{amssymb}
\usepackage{latexsym}
\usepackage{graphicx}
\usepackage[english]{babel}
\usepackage{multirow}
\usepackage{float}
\usepackage{url}
\usepackage{hyperref}
\usepackage{slashed}
\usepackage{xcolor} 
\usepackage{verbatim}

\hypersetup{colorlinks, citecolor=bluscuro, linkcolor=black, urlcolor=bluscuro}
\definecolor{rossos}{cmyk}{0,1,1,0.55}
\definecolor{bluscuro}{rgb}{0.15, 0.2, .85}
\definecolor{bluchiaro}{cmyk}{1,.3,0.,0.1}

\usepackage[normalem]{ulem}

\def\sfrac#1#2{{\textstyle{#1\over #2}}}
\newcommand{\be}{\begin{equation}}
\newcommand{\ee}{\end{equation}}
\newcommand{\ba}{\begin{array}}
\newcommand{\ea}{\end{array}}
\newcommand{\bea}{\begin{eqnarray}}
\newcommand{\eea}{\end{eqnarray}}

\newcommand{\nn}{\nonumber}

\renewcommand{\dh}{\delta_\lambda}

\def\simlt{\stackrel{<}{{}_\sim}}
\def\simgt{\stackrel{>}{{}_\sim}}
\def\bma#1{\mbox{\boldmath{$#1$}}}

\newcommand{\arXiv}[2]{\href{http://arxiv.org/pdf/#1}{{\tt [#2/#1]}}}
\newcommand{\arXivold}[1]{\href{http://arxiv.org/pdf/#1}{{\tt [#1]}}}

\begin{document}

\title{Axionic Landscape for Higgs Near-Criticality}

\author{James M.\ Cline}
\affiliation{McGill University, Department of Physics, 3600 University St.,
Montr\'eal, Qc H3A2T8 Canada}
\author{Jos\'e R.\ Espinosa}
\affiliation{
Institut de F\'\i sica d'Altes Energies (IFAE), The Barcelona Institute of Science and
Technology (BIST), Campus UAB, 08193, Bellaterra (Barcelona), Spain
and\\
ICREA, Instituci\'o Catalana de Recerca i Estudis Avan\c{c}ats, Pg.\ Llu\'\i s Companys 23, 08010 Barcelona, Spain}

\begin{abstract} The measured value of the Higgs quartic coupling
$\lambda$ is peculiarly close to the critical value above which the
Higgs potential becomes  unstable, when extrapolated to high scales by
renormalization group running.  It is tempting to speculate that there
is an anthropic reason behind this near-criticality.  We show how an
axionic field  can provide a landscape of vacuum states in which
$\lambda$
scans.  These states are populated during inflation to create a
multiverse with different quartic couplings, with a probability
distribution $P$ that can be computed.    If $P$ is peaked in the
anthropically forbidden region of Higgs instability, then the most
probable universe compatible with observers would be close to the
boundary, as observed.  We discuss three scenarios depending on the Higgs vacuum selection mechanism: decay by quantum tunneling; by thermal fluctuations or by inflationary fluctuations.
\end{abstract} \maketitle

\section{Introduction}

\begin{figure}[b]
\hspace{-0.4cm}
\centerline{
\includegraphics[width=0.92\hsize]{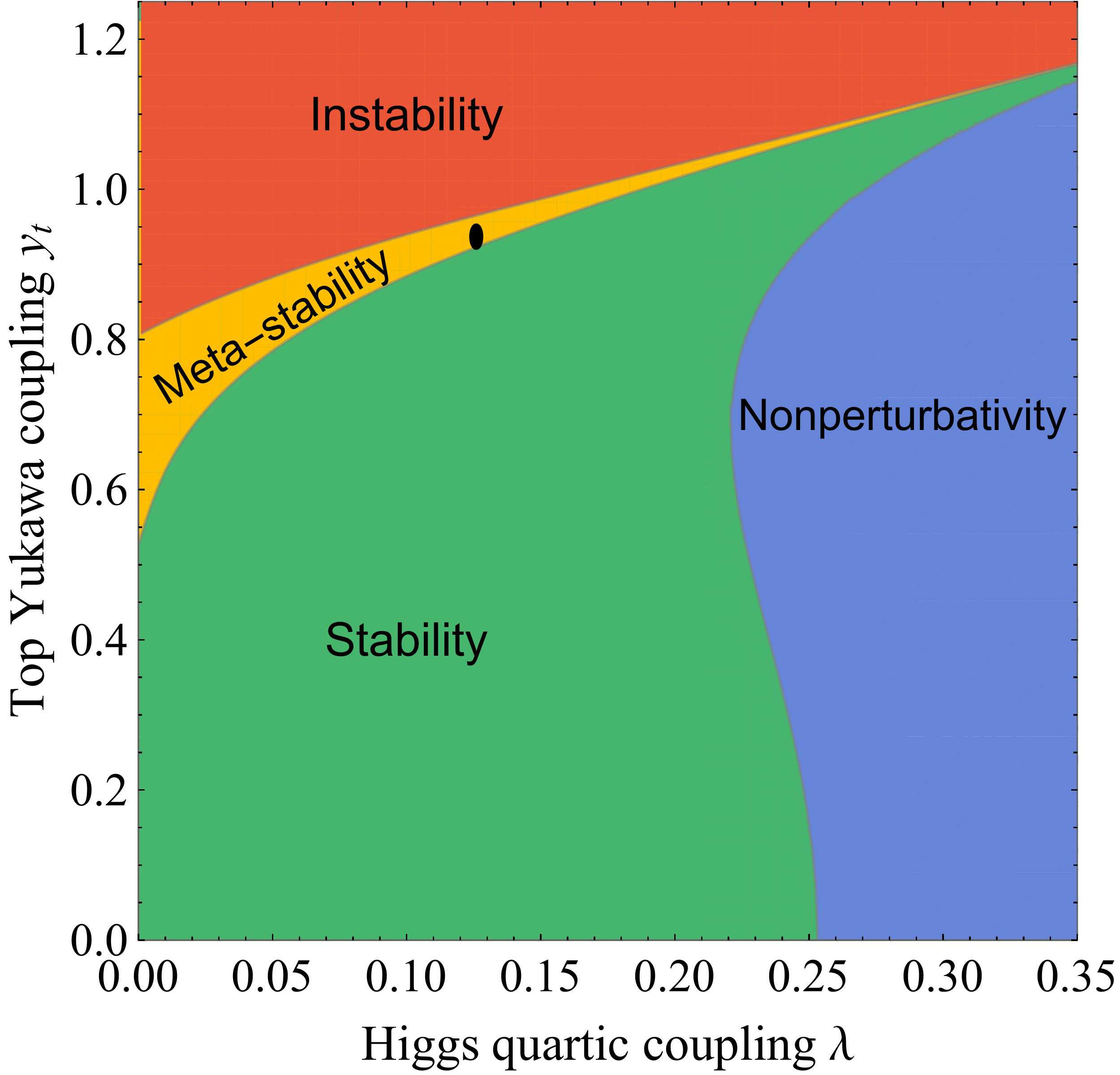}}
\caption{Regions of the $\lambda$-$y_t$ plane leading to
stability, metastability or instability of the Higgs potential at high
scales (at NNLO accuracy \cite{STAB}). In the region labeled ``Nonperturbativity'' $\lambda$ becomes strong below the Planck scale. The couplings are defined at the electroweak scale.}
\label{fig:stability}
\end{figure}

The standard model (SM) of particle physics, while enjoying tremendous
success as an accurate description of nature, has many parameters
whose values look mysterious from a theoretical perspective. Why are
the Higgs mass and the energy scale of the cosmological constant so
small compared to the Planck scale?  Why is $\theta_{QCD}$ so small? 
What is the origin of the hierarchy of fermion masses? Such questions
have inspired many efforts to go beyond the standard model.  Following
the discovery of the Higgs boson,  there is a new item, dubbed ``Higgs
near-criticality,'' on the list: why is the Higgs self-coupling
$\lambda$  (in conjunction with the top quark Yukawa coupling $y_t$) 
so close to   the critical value beyond which the Higgs potential
becomes unstable at high scales?  The situation is illustrated in
fig.\ \ref{fig:stability} \cite{STAB}, which shows the
regions of stability, metastability and instability of our vacuum, in
the $\lambda$-$y_t$ plane, with the small ellipse of the measured
values falling in the narrow region of metastability. In the
metastability (instability) region the vacuum lifetime is longer
(shorter) than the age of the Universe.

The answer could of course be that it is a coincidence: for fixed
$y_t$, the quartic coupling is $0.01$ below the stability boundary
($0.03$ above the instability line),
which is a tuning of only 8\% (23\%) relative to its actual value.  On the
other hand if $\lambda$ could {\it a priori} have taken any value between
zero and $4\pi$, this becomes a tuning of $0.08\%$ ($0.2\%$),  more in accord with
the visual impression from fig.~\ref{fig:stability}.  This is
predicated on the assumption that there is no new physics coupled to the Higgs at high scales (up to the Planck scale) that might shift the stability boundaries relative to where they are shown.   Nevertheless since there is an
anthropic reason for $\lambda$ to avoid the instability region, it
is tempting to construct a scenario where this explains the
coincidence.

While anthropic reasoning is eschewed by many physicists, if
there is a landscape of vacuum states in which anthropically sensitive
parameters are sampled, it seems difficult to dismiss.  For example
the very large number of flux compactifications in string theory
\cite{Douglas:2003um,Ashok:2003gk} make it plausible that our universe
is part of a much larger multiverse \cite{LindeEternal}.   A solution of the cosmological 
constant ($\Lambda$) problem was proposed in which $\Lambda$ is
finely scanned by these flux vacua \cite{BP}, yielding 
values consistent with anthropic bounds \cite{Weinberg:1987dv}.
Coleman's wormhole mechanism \cite{Coleman:1988tj} is another example
of a multiverse in which the most likely value of $\Lambda$ is small
(in fact vanishing). 

In this context, Rubakov and Shaposhnikov argued \cite{Rubakov:1989pn}
that the observed values of physical constants might generically  be
close to the boundaries of the anthropically allowed regions.  If the
probability distribution is such that the most likely value
of a parameter is anthropically forbidden, then the most likely
observed value would be close to the boundary, since there are no
observers on the forbidden side.  The near-criticality of the Higgs
potential looks like a possible example of this phenomenon.

The anthropic necessity of Higgs stability is an old observation
that was used to put a lower bound on the Higgs mass (or an upper
bound on the heaviest quark mass) as early 
as 1979 \cite{Stab0}.  Improved predictions using
higher orders in the loop expansion were subsequently made
\cite{STAB,StabMore}.   An indication of 
how delicate the tuning is for near-criticality is provided by 
the comparison of such predictions at different levels of precision \cite{ELattice}:
at LO our vacuum would be deep in the instability region, at NLO
in the middle of the metastable one and at NNLO very close to
the stability boundary.

Of particular relevance for our work, the implications of Higgs
stability within a landscape of vacua with $\lambda$ scanning were
studied in ref.\  \cite{Hall}, assuming conditions just like those
suggested by ref.\ \cite{Rubakov:1989pn} for the underlying
probability distribution $P(\lambda)$, namely that it is maximized
in the unstable region of small $\lambda$.  In that work, a
model-independent analysis was done, where no particular model of 
the landscape was proposed; rather a reasonable functional form for
$P(\lambda)$ was assumed, which led to predictions for the Higgs mass prior
to its measurement.  

We think it is worthwhile to revisit the question of Higgs
near-criticality within a specific model of the landscape, since
such a study may reveal nontrivial challenges to the overall
consistency of such a picture, that may be shared by other 
possible examples.  At the same time we introduce a new kind of
landscape that is particularly simple and amenable to calculations,
namely the vacuum states provided by the minima of the potential of an axion field (whose detailed properties are very
different from those of the QCD axion).

We are inspired  by a string-theory-motivated construction,  axion
monodromy, previously  used for inflation
\cite{AxMono} and by the  relaxion mechanism
\cite{relaxion}, used for solving the weak scale hierarchy problem.  In
contrast to these applications however, we wish to avoid classical
evolution of the axion $a$ during cosmological evolution.  Instead,
the universe is assumed to split into causally disconnected domains
where $a$ sits in different local minima of its potential.  These 
vacuum states were populated by quantum fluctuations of $a$ during a period of  inflation, are essentially stable against tunneling once formed, and
so realize a tractable example of a multiverse.  The probability distribution
is calculable in terms of the axion potential, given certain
assumptions about the cosmological evolution that we will specify.

\section{Landscape for $\bma{\lambda}$}

The field $a$ has a potential of the form
\bea
	V(a) &=&  \overline V(a)
-  \Lambda_b^4\cos(a/f) \ ,
\label{Va}
\eea
where $\overline V$ denotes the part of the potential that can be approximated as non-oscillatory on a field range large compared to $2\pi f$.
As the field $a$ has an axionic origin  ($a$ is a pseudo-Goldstone
boson, like a phase field), it originally enjoys a shift
symmetry $a\rightarrow a +c$ that is broken by the potential (\ref{Va}).
The term $\Lambda_b^4\cos(a/f) $ breaks the shift symmetry  down to a discrete subgroup $a\rightarrow a +2\pi f$, while $\overline V(a)$ breaks the
shift symmetry completely (at least in the range we consider; see below). It is then natural to expect these breaking terms to be
much smaller than the typical mass scale or cutoff of the theory that
 we will call 
$\Lambda$. In a string theoretic UV completion, $\Lambda$ could be
the string scale.

We assume then $\Lambda_b \lesssim \Lambda$ and take 
$\overline V(a) = \Lambda^4\ V(\eta a /\Lambda)$, with $\eta\ll 1$. 
For our purposes
it will suffice to keep the linear term of this function:
\be
\overline V(a) = \eta \Lambda^3 a +\dots
\label{smoothV}
\ee
This linear term should accurately describe 
the potential $\overline V(a)$
in a typical field region.  Without loss of generality (by doing 
a shift in the field), we can 
take this typical region to be in the vicinity of $a=0$ and
we can also take $\eta>0$ so that $\overline V(a)$ is a growing function of $a$.

A concrete example for $\overline V$ that arises in certain string theory
compactifications \cite{AxMono} is
\be
	\overline V(a) = M^4\sqrt{1 + a^2/F^2}\simeq   M^4 a/ F,
\label{monod}
\ee
where the  linear approximation is valid  in the region where
$a\gg F$. Here, $M$ and $F$ are generically  at the string scale, but
if the axion arises from a warped throat, then $M$ can be
parametrically suppressed by a  warp factor, which may be
exponentially small. 

Another example is the clockwork axion \cite{clockwork}, with $\overline V(a)=\epsilon \Lambda^4 \cos(a/F)$, and $F=N f \gg f$, 
which hierarchy can be arranged in a natural way. In this setting, the field range is compact, $2\pi F$, but we are interested in a patch $\Delta a$ with $2\pi f \ll \Delta a \ll 2\pi F$, and there
we can expand $\overline V(a)$ as in (\ref{smoothV}) around some
typical value $a_0$, obtaining $\eta = -(\epsilon \Lambda/F) \sin(a_0/F)$.

Let the minima of the potential (\ref{Va}) be labeled by an integer $n$, 
such that $a_n \simeq 2\pi n f$. 
A basic condition for having a landscape is that 
$\overline V$ must be sufficiently flat 
so that it does not destroy the local minima of the oscillatory
part.  This requires 
\be
	\overline V{}'(a) = 
	 \eta\,\Lambda^3
	\lesssim {\Lambda_b^4\over f}\ ,
\label{cond1}
\ee
which, if satisfied, would naively imply infinitely many local minima.
In realistic string constructions however, there is back-reaction from
large windings, so that the actual number of minima is limited to
$N \lesssim 10-100$, beyond which the above description breaks down, and possibly an extra dimension decompactifies \cite{Liam}.
In clock-work constructions the number of vacua is also finite as the field range is compact.

We assume that in addition to $\overline V$, there is a coupling
of $a$  to the Higgs potential:
\be
	V_h = \Big(-\mu_h^2 + c_h \eta\, a \Lambda \Big) |H|^2 +
\left(\lambda+ c_\lambda\, \frac{\eta \, a}{\Lambda}\right) |H|^4\ .\label{Vah}
\ee
Such couplings also break the shift symmetry and so we assign a factor $\eta$ to them. 
The $a$-terms in (\ref{Vah}) could be regarded as arising from a generalization of eq.\ (\ref{monod})
by taking $M^4 \to M^4 + {\cal O}(\Lambda^2|H|^2, |H|^4)$ or
from expanded $\cos(a/F)$ potentials in the clockwork realization.
In the landscape of vacua of the $a$ field, 
where $\langle a\rangle =
a_n \simeq 2\pi n f$, this shifts the bare values ({\it i.e.,} the values at the UV scale $\Lambda$) of the Higgs 
parameters to
\bea
\label{scanning}
	\mu_n^2 &=& \mu_h^2 - n\, c_h \eta\,(2\pi f)\Lambda\,,\nn\\ 
 	\lambda_n &=& \lambda +  n\,  c_\lambda\eta \frac{2\pi
f}{\Lambda}
	\equiv \lambda + n\,\delta\lambda\, .
\label{dleq}
\eea
Here we assume that some other mechanism solves the
weak scale hierarchy problem ({\it e.g.} a relaxion mechanism \cite{relaxion}) so that $\mu_n$ is of electroweak size and focus on the shift in the Higgs coupling. For reasons detailed below we also choose $c_\lambda>0$.  Likewise we must assume there is another
mechanism for solving the cosmological constant problem, since the
vacuum energy varies between $a$-vacua due to the nonperiodic part of the potential $\overline V$.  

\begin{figure}[t]
\hspace{-0.4cm}
\centerline{
\includegraphics[width=0.92\hsize]{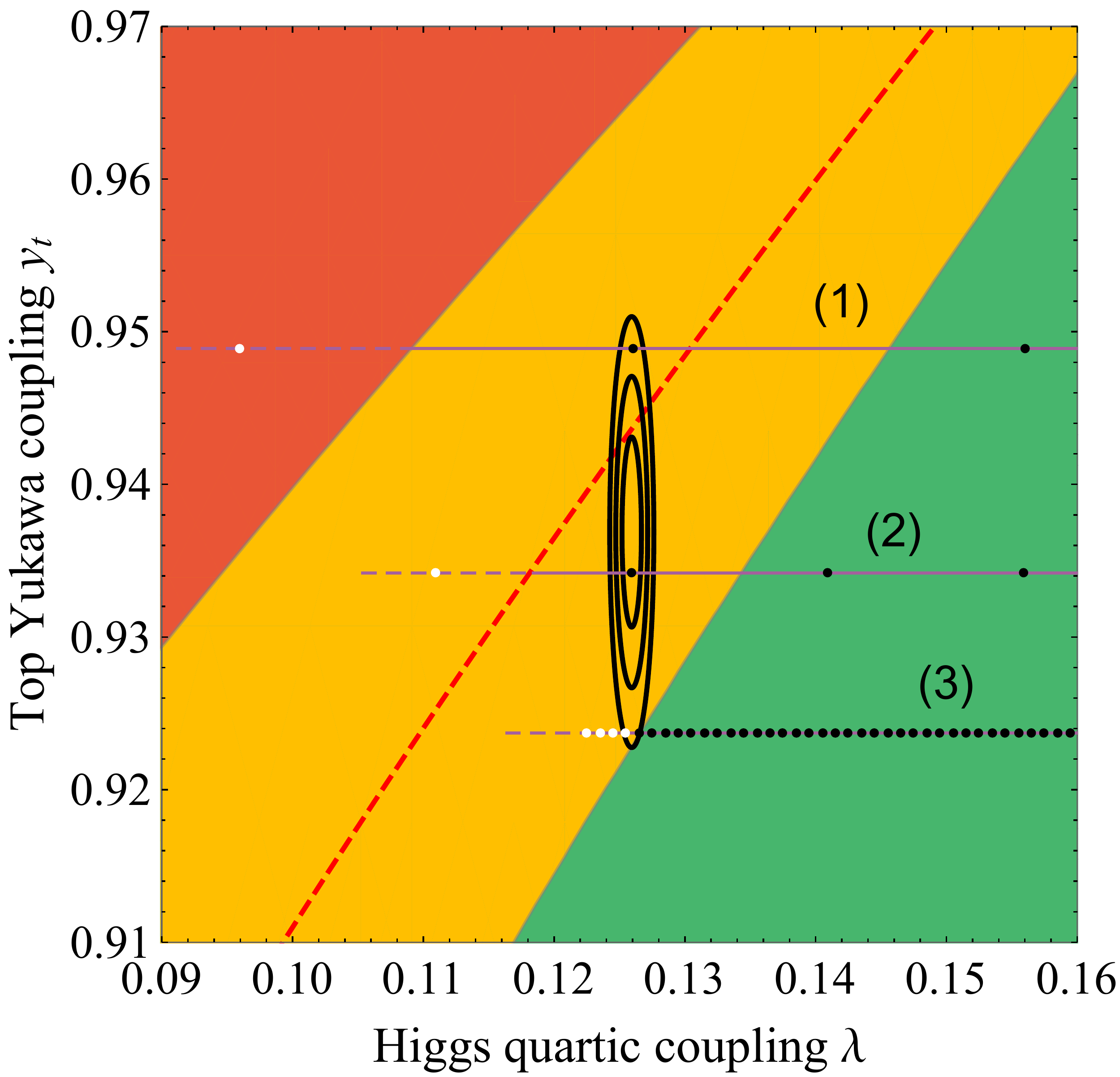}}
\caption{Zoom-in of Fig.~\ref{fig:stability} showing also the instability lines for thermal vacuum decay with $T_R=10^{14-16}$ GeV (red dashed lines). Trajectories of $a$-vacua are shown
(surviving ones in black, doomed ones in white) for the three cases discussed in the text. We use $M_h=125.09\pm 0.24$ GeV \cite{mh} and $M_t=173.34\pm 0.67$ GeV \cite{top} at 1-$\sigma$ for the experimental ellipses.}
\label{fig:zoom}
\end{figure}

We consider three possible scenarios, each associated to one of the
three critical boundaries shown in fig.\ \ref{fig:zoom}; these are
the boundaries of instability and metastability at zero temperature,
and the boundary of high-temperature instability that depends upon the 
assumed reheating temperature (dashed lines).  Our mechanism explains
why we would observe  $(\lambda,y_t)$ to be near (and to the right of) one of these boundaries.
The characteristics of the
three categories are summarized in table \ref{tab}. 
Fig.~\ref{fig:zoom} shows trajectories of successive vacua that
exemplify each case. Which one of the three is actually realized
depends upon cosmological parameters, as we will discuss
in more detail in the next section.

{\renewcommand{\arraystretch}{1.5} 
\begin{table}
\centering
\small
\begin{tabular}{| c |  c | c | c |}
\hline 
& (1)  & (2) &  (3) \\  \hline \hline 
Boundary & $T=0$ Instablity & $T_R$ Instability & Stability \\ \hline
Vacuum & Quantum & Thermal  & Inflationary  \vspace*{-3pt}\\ 
 Selection & $T=0$ decay &  decay &  decay \\ \hline
$\delta \lambda$ &  $\sim 0.05$  & $\sim 0.02$ & $\ll 0.01$  \\ \hline
$M_t$/GeV  & $173.34\pm 2.28$  & $ 173.34^{+1.34}_{-2.28}$ & $\simeq 171$ 
\\ \hline
\end{tabular}
\caption{\it Characteristics of the three cases we consider in the text, 
{regarding} the critical boundary, vacuum selection mechanism, step in $\lambda$ needed and range of top mass (inside the experimental $3\sigma$ band) required.}
\label{tab}
\end{table}
}

In case (1) we end close to the instability boundary and the probability to live in vacua beyond that boundary is depleted by $T=0$ decay, 
in which the Higgs vacuum has a lifetime that is shorter than the age of the Universe. To explain why a
point lying in the experimentally allowed ellipse at $y_t\simeq 0.95$
corresponds to the most probable anthropically allowed vacuum, we need $\delta\lambda\sim 0.05$, the approximate width of the metastable 
region.\footnote{This number can be estimated as follows.  The
vacuum decay rate per unit volume is $\Gamma\sim h_t^4
e^{-8\pi^2/(3|\lambda(h_t)|)}$, where $h_t$ is the preferred value for tunneling. The decay probability is $\Gamma$ times the 4D volume of our past light-cone 
$\sim (e^{140}/m_P)^4$. Decay probabilities of order one require
$\lambda(h_t)\sim -0.05$ and this number is confirmed by a more
sophisticated calculation (see {\it e.g.} \cite{EEGIRS}).  Thus the
metastable region is approximately $\lambda(h_t)\in\{-0.05,0\}$.  This
translates to the region shown in Fig.\ \ref{fig:stability} after running the
couplings down to the weak scale.}  
Scenario (1) could take place for any value of the top mass, within the
experimentally preferred region, which we take to be the 3$\sigma$
range $M_t=173.34\pm 3\times 0.67$ GeV \cite{top}. 

In case (2) we end in a vacuum near the instability boundary for decay by thermal fluctuations with a high reheating temperature,
that reduce the region of metastability. As concrete examples
we illustrate the cases of $T_R=10^{14}$
and $T_R=10^{16}$ GeV. The boundary of the reduced region is
shown as the dashed lines in  Fig.~\ref{fig:zoom} (see refs.\
\cite{Tdec,Higgstory}), and a possible trajectory illustrating this 
case is shown along $y_t\simeq 0.934$.  A smaller step size
$\delta\lambda \sim 0.02$ is suggested for naturally explaining the
distance of the SM point from the dashed boundary.   This mechanism,
for such large $T_{\rm R}$ favors the lower range of the top mass,
with $M_t\simeq 173.34^{+1.34}_{-2.28}$ GeV.

In case (3) we end very close to the stability boundary beyond which
the Higgs vacuum is unstable against decay during inflation, for
sufficiently large values of ${\cal H}_I\sqrt{N_e}$. This case is
illustrated by the trajectory passing through the bottom of the
experimental ellipse.  Here the most probable state would be
the one closest to the boundary in the absolute stability region, and
it would require a very small step size $\delta\lambda$ to be
naturally close to the  experimental ellipse.  Although this
possibility is currently disfavored,  it is not excluded and provides
another possible regime for explaining near-critical
stability, if the top mass is very close to its lowest
$3\sigma$ value, $M_t\simeq 171$ GeV.

Once $\delta\lambda$ is fixed, (\ref{dleq}) can be used to 
to eliminate the unknown parameter $\eta$ in terms of $f$ and
$\delta\lambda$. We introduce the ratio $\dh \equiv
\delta\lambda/0.05$ [which is of order unity in case (1)] to allow for the
possibility of any of the three cases.  Hence
\be
c_\lambda \eta = 0.05\,\delta_\lambda\, \frac{\Lambda}{2\pi f}\ .
\label{eta}
\ee

\section{Probability distribution of vacua}
\label{prob_sec}

A key ingredient of our scenario is the process by which
the vacua get populated by quantum fluctuations during inflation, and
the resulting probability distribution function $P(t,a_n)$ for the different
vacua.  It is governed by the Fokker-Planck
equation
\be
	{\partial P\over dt} = {\partial\over\partial a}
	\left[{V'(a)\over 3 {\cal H}_I}P + {{\cal H}_I^3\over 8\pi^2}{\partial P\over\partial a}
	\right]\ ,
\label{FPeq}
\ee
(see for example refs.\ \cite{Linde,EGR,Zurek}) where ${\cal H}_I$ is the Hubble
parameter during inflation.  We take the inflationary
contribution to the energy density to be much larger than $V(a)$ and 
consider ${\cal H}_I$ to be approximately constant.  
Then the 
stationary solution to (\ref{FPeq}) is\footnote{If $a$ contributes significantly
to the energy density, the stationary solution is 
$P(a) \sim \exp \{ 24\pi^2 m_P^4/[V_I+V(a)]\}$, 
where $V_I$ is the inflaton field potential and $m_P$ the
reduced Planck mass. 
An expansion for small $V(a)/V_I$ reproduces Eq.~(\ref{Peq}).} 
\be
	P(a) \sim e^{ -8\pi^2 V(a)/3 {\cal H}_I^4}\ .
\label{Peq}
\ee
We  assume for the moment that this
stationary solution  (\ref{Peq}) is reached and determines the relative probabilities of the different vacua (disregarding for now the possible decays along the Higgs direction).  The necessary
conditions to justify this assumption will be discussed below. We do not care about the 
normalization of $P(a)$ as we are only interested in relative probabilities between different vacua.

At the local minima of the potential we have $V(a_n)\simeq \overline V(a_n)$,
neglecting the uninteresting constant contribution $-\Lambda_b^4$ and
taking $v\ll \Lambda$, where $v=246$ GeV is the Higgs vacuum
expectation value, with $v^2/2=\langle |H|^2\rangle$.  With our
convention $ \eta>0$, the underlying landscape probability
distribution prefers the more negative values of $n$, which reduce
$\overline V(a_n)$. By choosing $c_\lambda\eta>0$ we then favor
negative $\lambda_n$ in eq.\ (\ref{scanning})  and unstable Higgs
potentials are preferred within the landscape.

In order to have significant variation of $P(a_n)$ 
near the instability boundary, the exponent of (\ref{Peq}) should 
change by $O(1)$ between neighboring vacua.  The ratio of the
probabilities of  the second and first anthropically allowed vacua, relative to
the anthropic boundary, is given by 
\be
	-\ln {P_2\over P_1} =  {8\pi^2\Delta\overline V\over 3
{\cal H}_I^4} \ \simgt\  {\cal O}(1) \ ,
\label{dPeq}
\ee
where 
\be
	\Delta \overline V =  \eta (2\pi f) \Lambda^3 = 0.05\, 
	\Lambda_r^4\ .
\label{dveq}
\ee
In the last step we removed $\eta$ by using (\ref{eta})
and introduce the quantity $\Lambda_r$ (that appears repeatedly) as
\be
\Lambda_r\equiv \left({\dh}/{c_\lambda}\right)^{1/4}\Lambda\  .
\ee

\begin{figure}[t]
\hspace{-0.4cm}
\centerline{
\includegraphics[width=0.92\hsize]{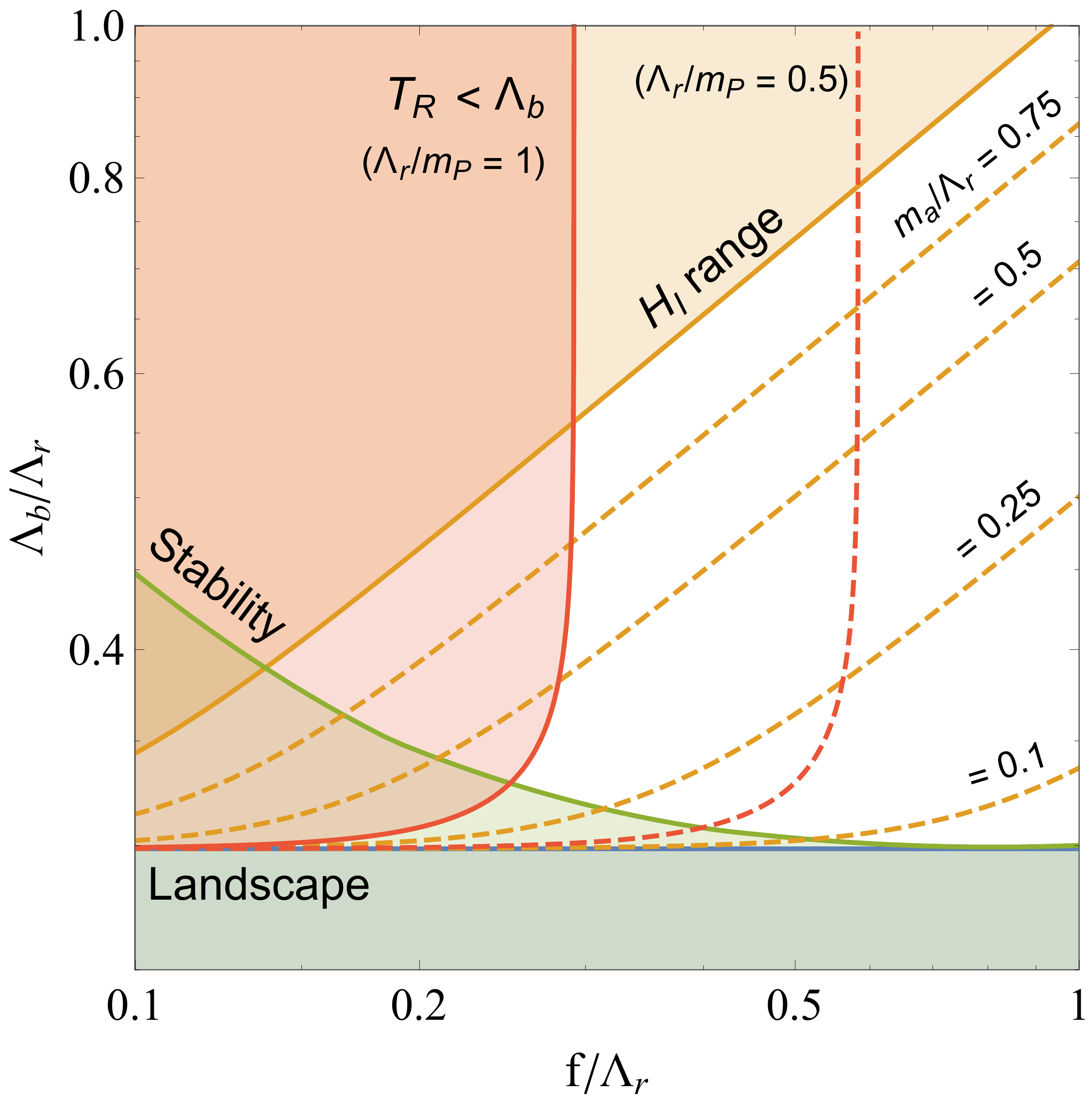}}
\caption{Excluded (shaded)  regions in the
plane of $\Lambda_b/\Lambda_r$
 versus $f/\Lambda_r$.  ``Landscape'' region violates 
condition (\ref{cond1}); ``Long Inflation'' region violates
condition (\ref{noeternal}); ``Stability'' denotes the vacuum 
stability bound (\ref{T0rate}); ``$H_I$ range'' curve
denotes the limit beyond which the interval of allowed 
inflationary Hubble
rates ${\cal H}_I$ from  (\ref{HIrange}) vanishes.
The dashed diagonal lines indicate the lowest allowed value of
${\cal H}_I/\Lambda_r$ in this allowed range
as the axion mass is varied (see text).   The bounds corresponding to
``$T_R < \Lambda_b$'' depend upon $\Lambda_r/m_P$ and are shown for
two values of that ratio. 
 }
\label{fig:bounds}
\end{figure}

Condition (\ref{dPeq}) will lead to the most likely anthropically allowed vacuum being 
the one closest to the critical boundary in question.  It
imposes a maximum value of the Hubble rate during inflation:
${\cal H}_I^4 \lesssim (8\pi^2\Delta\overline V/ 3)$. 
On the other hand, the derivation of the Fokker-Planck equation 
from the stochastic approach to tunneling \cite{Linde}
assumes that ${\cal H}_I > m_a$, the mass of the $a$ 
field. It is possible that this is only a sufficient and not
a necessary condition \cite{Linde2}, but if we respect it 
[along with (\ref{dPeq})] then
${\cal H}_I$ should be in the interval\footnote{Here we account for
the displacement away from the minimum of the cosine potential due to
the linear term, using $V'=0$ to eliminate $\cos(a/f)$ in $m_a^2 =
V''$, and (\ref{dveq}) to reexpress $\Delta \overline V$.}
\be
m_a=\frac{\Lambda_b^2}{f}\left[1-\left(0.3\frac{\Lambda_r}{\Lambda_b}\right)^8\right]^{1/4} \lesssim {\cal H}_I 
\lesssim 1.07\,\Lambda_r\ .
\label{HIrange}
\ee
The upper limit is plotted in Figure~\ref{fig:bounds} with the 
label ``$H_I$ range.''\ \ 
Information on the lower limit, which varies from point to point 
in the plane, is
conveyed by the dashed lines; {\it e.g.,} on the line labeled
``$m_a/\Lambda_r=0.25$,'' the interval for ${\cal H}_I/\Lambda_r$ is
(0.25,\,1.07).  On the other hand, Eq.\ (\ref{cond1}), required to
guarantee the existence of a landscape of $a$-vacua (which coincides with
the requirement $m_a>0$), gives the limit 
\be
\Lambda_b > 0.3\Lambda_r\, ,
\label{Lbrmin}
\ee
which is also plotted in  Figure~\ref{fig:bounds} and labeled 
``Landscape.''

If we also insist that the inflaton potential dominates
over the $a$ potential, then ${\cal
H}^2_I\gtrsim 2\pi\eta\Lambda^3 N f/3 m_P^2$, where we have assumed
that $a = 2\pi N f$ in the vicinity of our $a$-vacuum.  
Using (\ref{dveq}) to eliminate $\eta f$
and combining with the upper limit
in (\ref{HIrange}) we find
\be
\label{Lambda_lim}
\frac{\Lambda_r}{m_P}\simlt \frac{8.4}{\sqrt{N}} \ ,
\ee
which is not very constraining ({\it e.g.} if $N\lesssim 100$ or {$\Lambda_r\ll \Lambda$}).

\section{Vacuum stability}
\label{vac_stab}

For our own $a$-vacuum to be habitable, it must not 
decay too quickly through tunneling to neighboring axionic vacua
(not to be confused with the possible {decay along the Higgs direction}). 
This might occur during inflation, after reheating, when 
the effect of finite temperature is important, or at late times
when we can consider $T$ to be zero. 

At zero temperature, the criterion for  vacuum stability becomes
\be
	A e^{-S_4} \lesssim {\cal H}_0^4
\label{T0rate}
\ee	
where ${\cal H}_0$ is the present Hubble constant ($\sim e^{-140}m_{P}$ in Planckian units). $S_4$ is the 4D Euclidean action for critical bubbles corresponding to
transitions between neighboring vacua \cite{Coleman:1977py}.
In (\ref{T0rate}), the prefactor $A = (S_4/2\pi)^2 J$, with $J$ being a ratio of functional
determinants with dimensions of [mass]$^4$.  
The $J$ factor is
difficult to compute, but is expected to be of order $\Lambda_b^4$
 or $f^4$, always smaller than $\Lambda^4$ and $m_{P}^4$,
 so it is conservative to require $S_4\gtrsim 560$ as a condition for vacuum stability. 
We numerically compute the bounce solution and resulting $S_4$ and
plot this stability condition,  labeled ``Stability,'' in Figure~\ref{fig:bounds}.

\begin{figure}[t]
\hspace{-0.4cm}
\centerline{
\includegraphics[width=0.92\hsize]{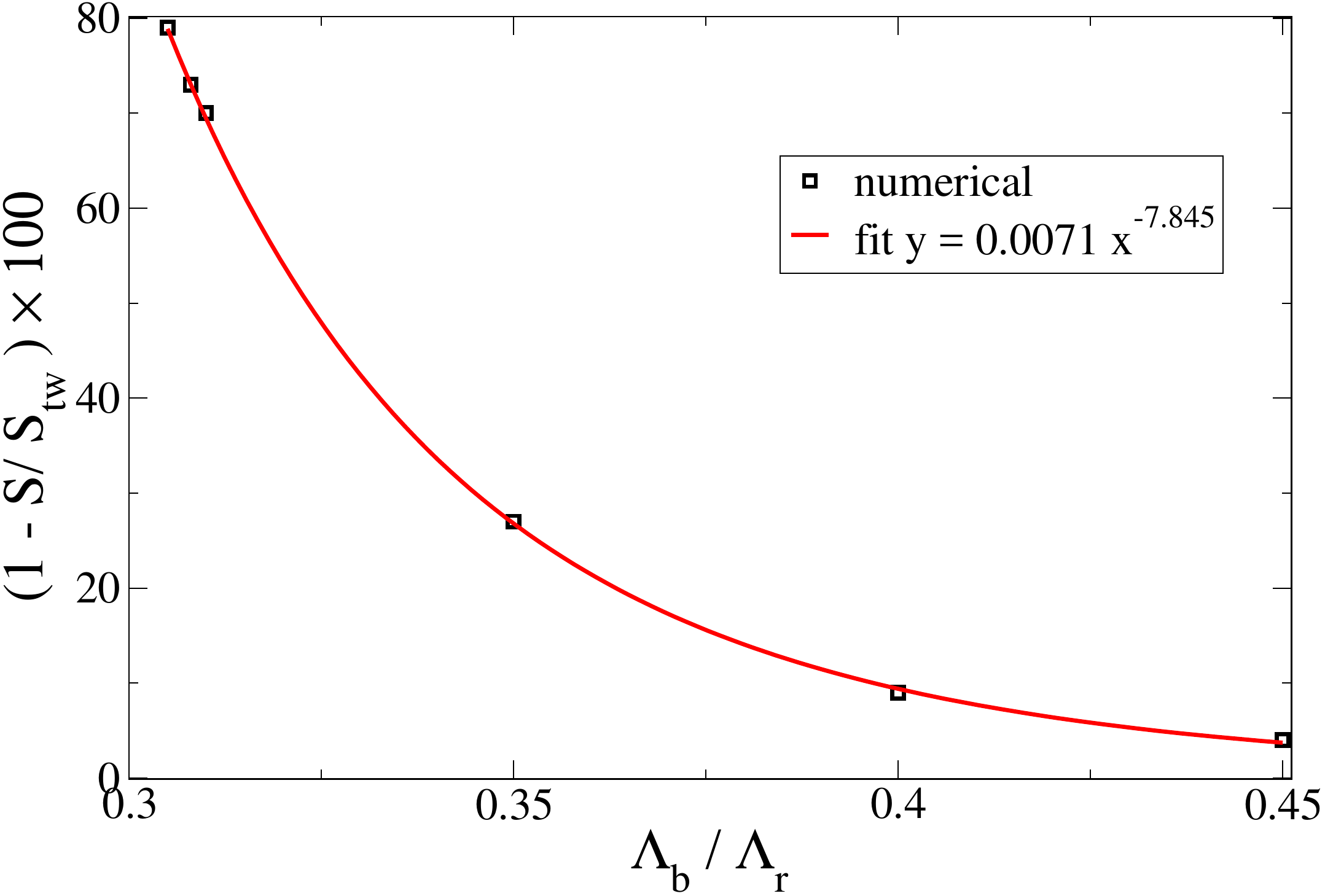}}
\caption{Fractional error in the thin-wall approximation 
for the 4D tunneling action, as a function of $\Lambda_b/\Lambda_r$.
The analytic fit (\ref{twfit}) as well as the numerical results are
shown.}
\label{fig:tw}
\end{figure}

An analytic formulation of the stability criterion can be obtained
using the thin-wall approximation \cite{Coleman:1977py}, in which the
4D action is 
\be
	S_{4,tw} \simeq {27\pi^2\over 2}\, {\sigma^4\over \Delta V^3}
	\ ,
\label{S4eq}
\ee
depending upon the bubble wall tension 
\be
	\sigma \simeq \int_{0}^{2\pi f} da\,
	\sqrt{2 \Lambda_b^4[1-\cos(a/f)]} = 8 \Lambda_b^2 f\ ,
\ee
and the potential difference between neighboring vacua as 
given by (\ref{dveq}).  By numerical calculation of the actual
tunneling action, we find that this approximation is not very good
in the region of parameter space of interest; however by comparing the
exact and approximate results it is possible to correct for this.
The relevant parameter determining how well the thin-wall
approximation works is $\Lambda_b/\Lambda_r$,\footnote{{By the rescalings $\hat a = a/f$ and $x =
r \Lambda_b^2/f$, we can write
$S_4 = 2\pi^2 (f/\Lambda)^4\int dx\,x^3\left[\sfrac12 \hat a'^2 +
(0.3\Lambda_r/\Lambda_b)^4\hat a - \cos \hat a\right]$, using
(\ref{dveq}).  The thin-wall approximation breaks down as the 
coefficient of the linear term becomes large.}}\ \  and we find
that the fractional error in the action can be accurately fit to the formula
\be
	1-{S_4\over S_{4,tw}}   \simeq 7.1\times
10^{-5}\left(\Lambda_b\over\Lambda_r\right)^{-7.845}
\label{twfit}
\ee
where $S_4$ is the full numerical value.  This function is shown in
fig.\ \ref{fig:tw}.

In the case of vacuum transitions due to thermal excitation  over the
barrier, one should estimate the 3D action for critical bubbles,
taking also into account the thermal corrections to the potential.
This is not  a straightforward task: it depends on possible couplings
of $a$ to other sectors of the theory and is limited to temperatures
well below the critical temperature  $T_c$ above which the dynamics
responsible for the nonperturbative generation of the barriers in the
axion potential become ineffective, but this is unspecified in our scenario.
If the reheating temperature $T_R$ is above $T_c$ one expects the
effective temperature-dependent barrier height 
$\Lambda_b^4(T)$ to start falling as a power of $T$ \cite{Preskill:1982cy}.
Given the level of uncertainty on
$T_c$, we content ourselves with imposing the condition that $T_R <
T_c\sim \Lambda_b$, as a
rough estimate for $T_c$.

To obtain $T_R$ we use the relation for the Hubble parameter 
during radiation domination ${\cal H}_R = 0.33\,\sqrt{g_*}\, 
T_R^2/m_P$. Assuming instant reheating we have 
${\cal H}_R ={\cal H}_I$ with ${\cal H}_I$ respecting 
 (\ref{HIrange}), which translates into the range
\be
0.54 \frac{\sqrt{m_a m_P}}{\Lambda_r}
 <\frac{T_R}{\Lambda_r}  \left(\frac{g_*}{g_*^{\rm SM}}\right)^{1/4}< 0.56 \sqrt{\frac{m_P}{\Lambda_r}}\ , 
\ee
with $g_*^{\rm SM}=106.75$. We exclude a point in parameter space if the lower limit of this range is bigger than $\Lambda_b/\Lambda_r$.
The resulting limit is shown in Fig.~\ref{fig:bounds}, labeled $T_R<\Lambda_b$, 
for two representative values of $\Lambda_r/m_P =0.5,\,1$.

In cases (1) and (2) we must also consider the possibility of vacuum
decay along the Higgs field direction, since we end up in the
metastable region with respect to such decays.  Metastability here
means that quantum fluctuations at zero temperature are slow on the
time scale ${\cal H}_0^{-1}$, and it does not take  into account the
possibility that tunneling was triggered at an  earlier time by
inflation.   In fact during inflation, if $\cal{H}_I$ is higher than
the instability scale, the Higgs field can be pushed over the barrier
that separates the electroweak vacuum from the unstable region of
field space \cite{EGR,Zurek,Higgstory}, and this leads to an
upper bound on ${\cal H}_I\sqrt{N_e}$, where $N_e$ is the number of
e-folds.  As discussed in the next section, this kind of bound
can be  generically violated in our framework if a very long period of 
inflation is needed to guarantee that the stationary solution to the 
Fokker-Planck equation is reached. In fact, this is the vacuum
selection mechanism in case (3).

For cases (1) and (2) we then have to forbid such decays during inflation. A simple way of circumventing this danger is to have a nonminimal coupling $\xi |H|^2R$ between the
Higgs field and the Ricci scalar $R$ \cite{EGR}. During inflation,
$R=-12 {\cal H}_I^2$, and this provides a contribution
$12 \xi {\cal H}_I^2$ to the squared Higgs mass, that stabilizes the potential or suppresses Higgs fluctuations altogether (for $\xi>3/16$), relaxing the
bound on ${\cal H}_I\sqrt{N_e}$ \cite{Higgstory}. Subsequent to
inflation, during preheating the induced Higgs mass term oscillates
along  with the inflaton, and this can cause parametric  resonant
production of Higgses, whose associated classical field can probe the
instability region again \cite{preheatdec,Dani} and  trigger vacuum
decay. To avoid this, it is sufficient to have $\xi$ in the range
$(0.06-4)$ \cite{Dani}, which we assume to be the case for scenarios (1) and (2).

\section{Initial conditions}
We have assumed that the stationary solution of the Fokker-Planck
equation was achieved during inflation.  Here we consider
how long a period of inflation would be required to achieve this, starting from some
different initial condition, for example that $P(a)$ was peaked around
the true vacuum state.  The barriers between neighboring vacua
must be large enough to prevent tunneling at late times, 
while the scale of 
inflation must be sufficiently low so that 
$P(a)$ is not too flat, eq.\ (\ref{dPeq}).  Both of these tend to
slow  the
time evolution of $P$.

It is instructive to consider a toy model consisting of a double-well
potential $V(\phi)$ with just two vacuum states, separated by a
barrier height $V_b$ that is large compared to the energy difference
between the two vacua.  The system is initially 
sharply
localized in one of the vacua, $\phi_1$, and allowed to evolve in
time according to the
Fokker-Planck equation.  By a combination of numerical
and analytical methods one discovers two relevant time scales,
hierarchically different. The shorter one, $\tau_1\simeq 3{\cal
H}_I/[2V''(\phi_1)]$, is associated with the
spread of $P$ until it reaches an approximately Gaussian shape around
the starting vacuum, $P(\phi)\simeq
\exp[-(\phi-\phi_1)^2/(2\sigma_1^2)]$, with $\sigma_1^2=3 {\cal
H}_I^4/[8\pi^2V''(\phi_1)]$. This solution is valid for small displacements and
is quasi-stationary. The long time scale, $\tau_t$, is associated with
the probability leakage to the second vacuum at $\phi_2$, through the
top of the barrier, at $\phi_t$. The associated rate,
$\Gamma=1/\tau_t$, is 
\be
	\Gamma \sim {{\cal H}_I^3\over 16 \pi^2 \sigma_1 \sigma_t}\,
	e^{-8\pi^2 V_b/3 {\cal H}_I^4}\ ,
\ee
where $\sigma_t^2 =  3 {\cal H}_I^4/(8\pi^2 |V''(\phi_t)|)$.

Applying this estimate to our scenario, we see that 
to avoid an exponentially long period of inflation, one
needs ${\cal H}_I^4 \gtrsim 8\pi^2 V_b/3$, while 
condition (\ref{dPeq}) 
implies ${\cal H}_I^4 \lesssim 8\pi^2\Delta\overline V/3$. Using
$V_b =\Lambda_b^4$ and $\Delta\overline V$ from (\ref{dveq}), 
the combined conditions
require
\be
\Lambda_b/\Lambda_r< 0.47\ .
\label{noeternal}
\ee
Hence it is possible to
satisfy all the criteria without having a very long period of inflation.

However, a more generic situation is to admit a prior period of
eternal inflation, which would automatically justify the stationary
solution since then an arbitrarily long period of evolution 
could occur prior to the  final stage of observable inflation.  Two common situations
can admit eternal inflation.  First, inflation could be chaotic during
the primordial stage, with the inflaton displaced high enough on its
potential so that upward quantum fluctuations can dominate over the
classical downhill evolution \cite{LindeEternal}.  Second, the inflaton (not necessarily
the same inflaton that is responsible for the final stage of inflation) could be
trapped in a false vacuum with an exponentially long lifetime, the
exponential of the tunneling action
\cite{LinVil}.  Either case allows us to relax
the requirement (\ref{noeternal}).

\section{Summary and conclusions}

We have presented a concrete realization of a mechanism to explain the
near-criticality of the SM Higgs quartic coupling $\lambda$. It uses
an axion-like field $a$ with a potential that develops a large number
of non-degenerate vacua in which $\lambda$ takes different values,
effectively scanning, due to a coupling of the Higgs to $a$. The vacua
are assumed to be populated during inflation with probabilities that
depend exponentially on the ratio  $V(a)/{\cal H}_I^4$. By 
appropriately choosing the sign of the overall slope of $V(a)$, vacua
with increasingly negative values of $\lambda$ are favored.  The
conditional probability for a particular vacuum state given that it is
compatible with observers, is zero if it undergoes catastrophic
decay of the Higgs vacuum.  Thus the most likely anthropically allowed
states are those that are close to a critical line in the plane of
$\lambda$ and $y_t$.  
 We discussed three different scenarios,
summarized in Table~1 and illustrated by Fig.~\ref{fig:zoom}.
They require different cosmological histories and parameters 
for the potential of the $a$ field,
and they depend upon the precise value of the top quark mass.

In case (1), vacua beyond the instability line are depleted by quantum
tunneling, which is faster than the age of the universe. In case (2),
that requires a large reheating temperature, thermal fluctuations over
the Higgs barrier  remove vacua beyond the thermal instability line.
In case (3), which  requires a high inflationary Hubble rate or a
large number of e-folds, Higgs fluctuations induced during inflation
trigger vacuum decay along the unstable Higgs direction, effectively
selecting vacua with stable Higgs potentials.

While the mechanism we have discussed offers an explanation for the
intriguing near-criticality of the Higgs quartic coupling, it does not
address the hierarchy problem. It would be quite interesting to  find
a mechanism that could address both issues simultaneously, especially
given the fact that similar mechanisms ({\it e.g.} relaxions) offer
potential solutions to the hierarchy problem.

It is perhaps disappointing that this scenario does not make
positive predictions for new physics at experimentally accessible
energies.
Since the only new field, the axion,
has a mass \{typically much larger than the electroweak scale, there are no manifestations at low
energy.  Instead, we predict an {\it absence} of new physics coupling to the
Higgs field at low scales, to the extent that such couplings would
move the critical lines of stability away from their standard model
values.  On the other hand, we think it is interesting that despite
the lack of low-energy experimental tests, the mechanism is highly
constrained by considerations of theoretical and cosmological
consistency.   It shows that the mere existence of a landscape is not
sufficient for a successful anthropic explanation of tuning problems.
Our results further indicate that the new physics scale should
generically be very high (not far below the string or Planck scale) to
make the vacua of the landscape stable against tunneling both during
inflation and at late times, and that a prior period of  eternal
inflation is strongly motivated.

{\bf Acknowledgments.}  J.M.C. thanks A.\ Linde, L.\ McAllister  and
M.\ Trott for helpful discussions, and the CERN Theory Department 
and Niels Bohr International Academy for hospitality
while this work was in progress, which was also supported by the Natural
Sciences and Engineering Research Council of Canada.  The work of J.R.E. has been partly
supported by the ERC grant 669668 -- NEO-NAT -- ERC-AdG-2014, the
Spanish Ministry MINECO under grants  2016-78022-P and
FPA2014-55613-P, the Severo Ochoa excellence program of MINECO (grant
SEV-2016-0588) and by the Generalitat grant 2014-SGR-1450.

\bibliographystyle{apsrev}

\end{document}